\documentclass[12pt]{article}

\usepackage{epsfig}
\input epsf

\title{Unintegrated gluon distributions in $D^{*\pm}$ and dijet associated 
photoproduction at HERA}

\author{A.V.~Lipatov, N.P.~Zotov}

\begin{document}

\maketitle

\begin{center}

{\it D.V.~Skobeltsyn Institute of Nuclear Physics,\\ 
M.V. Lomonosov Moscow State University,
\\119992 Moscow, Russia\/}\\[3mm]

\end{center}

\vspace{1cm}

\begin{center}

{\bf Abstract }

\end{center}

We consider the photoproduction of $D^{*\pm}$ mesons associated with 
two hadron jets at HERA collider in the framework of the $k_T$-factorization 
approach. The unintegrated gluon densities in a proton are obtained 
from the full CCFM, from unified BFKL-DGLAP evolution equations as
well as from the Kimber-Martin-Ryskin prescription. Resolved photon 
contributions are reproduced by the initial-state gluon radiation.
We investigate different production rates and make comparison with the
recent experimental data taken by the ZEUS collaboration.
Special attention is put on the specific dijet correlations 
which can 
provide unique information about non-collinear gluon evolution dynamics. 

\vspace{1cm}

\section{Introduction} \indent 

The production of heavy flavour (charm and bottom) in electron-proton collisions 
at HERA is a subject of intensive studies from both theoretical and 
experimental points of view~[1--5]. From the theoretical side, heavy quarks 
in $ep$ interactions can be produced via direct (photon-gluon fusion) and 
resolved production mechanisms. In resolved events, the photon emitted by 
the electron fluctuate into a hadronic state and a gluon and/or a quark of 
this hadronic fluctuation takes part in the hard interactions. 
It is expected that resolved photon processes contribute significantly in the 
photoproduction region, in which the photon is quasi-real ($Q^2 \sim 0$), 
and to be suppressed towards higher $Q^2$. Therefore charm and bottom 
photoproduction cross sections are sensitive to the parton (quark 
and gluon) content of the proton as well as of the photon. 

Usually quark and gluon densities are described by the 
Dokshitzer-Gribov-Lipatov-Altarelli-Parizi (DGLAP) evolution equation~[6] 
where large logarithmic terms proportional to $\ln \mu^2$ are taken 
into account only. The cross sections can be rewritten in terms of 
process-dependent hard matrix elements convoluted with quark or gluon 
density functions. In this way the dominant contributions come from diagrams 
where parton emissions in initial state are strongly ordered in virtuality. 
This is called collinear factorization, as the strong ordering means that 
the virtuality of the parton entering the hard scattering matrix elements can be 
neglected compared to the large scale $\mu$. However, at the high energies this hard 
scale is large compared to the $\Lambda_{\rm QCD}$ parameter but on the other 
hand $\mu$ is much smaller than the total energy $\sqrt s$ (around 300~GeV for 
the HERA collider). Therefore in such case it was expected that the DGLAP 
evolution, which is only valid at large $\mu^2$, should break down. The 
situation is classified as "semihard"~[7--10]. It is believed that 
at 
asymptotically large energies (or small $x \sim \mu^2/s$) the theoretically correct 
description is given by the Balitsky-Fadin-Kuraev-Lipatov (BFKL)
 evolution equation~[11] 
because here large terms proportional to $\ln 1/x$ are taken into account. 
Just as for DGLAP, in this way it is possible to factorize
an observable into a convolution of process-dependent hard matrix elements
with universal gluon distributions. But as the virtualities (and transverse 
momenta) of the propagating gluons are no longer ordered, the matrix 
elements have to be taken off-shell and the convolution made also over 
transverse momentum ${\mathbf k}_T$ with the unintegrated (i.e.~$k_T$-dependent) 
gluon distribution. This generalized factorization is called 
$k_T$-factorization~[7-10].

The unintegrated gluon distribution is a subject of intensive studies at present~[12, 13]. 
Various approaches to investigate this quantity have been proposed. 
So, there are unified BFKL-DGLAP equation~[14] which incorporate both the
resummed leading $\ln 1/x$ and the resummed leading $\ln \mu^2$ contributions.
Another approach, valid for both small and large $x$ also, have been developed by 
Ciafaloni, Catani, Fiorani and Marchesini, and is known as the CCFM model~[15]. 
It introduces angular ordering of emissions to correctly treat gluon coherence 
effects. In the limit of asymptotic energies, it is almost equivalent to BFKL~[16--18], 
but also similar to the DGLAP evolution for large $x$ and high $\mu^2$. The 
resulting unintegrated gluon distribution depends on two scales, the additional 
scale ${\bar q}$ is a variable related to the maximum angle allowed
in the emission and plays the role of the evolution scale $\mu$ in the
collinear parton densities. Also it is possible to obtain the two-scale
involved unintegrated gluon distributions from the conventional ones using
the Kimber-Martin-Ryskin (KMR) prescription~[19]. In this way the $\mu$ dependence 
in the unintegrated gluon distribution enters only in last step of the evolution, 
and single-scale evolution equations can be used up to this step.
The unintegrated gluon densities have the advantage that, in particular, they 
take into account true kinematics of the process under consideration even at leading 
order and are more suitable for less inclusive processes.

Recently the ZEUS collaboration has presented new experimental data~[2, 3]
on the charm production in electron-proton collisions at HERA,
namely the results of measurements of $D^{*\pm}$ meson 
production rates both inclusive and associated with one or two hadronic jets.
Concerning the theoretical treatment of charm photoproduction in the
framework of standard (collinear) QCD, two types of NLO calculations are 
available for comparison with the experimental data. The massive charm approach~[20] 
assumes that light quarks are the only active flavours in the structure functions of
the proton and photon, so that charm (and beauty) are produced only in the hard
process. In the massless scheme~[21, 22] charm and beauty are treated as an 
additional active flavours (massless partons). These two approach are applicable
in different regions: for $p_T^2 \simeq m_c^2$ and $p_T^2 \ge  
m_c^2$ accordingly. The massless charm calculations take into account charm 
excitation processes and thus predict a larger resolved component in comparison 
with the massive calculations. The photoproduction of a $D^*$ meson in association 
with a hadron jet was described recently in the next-to-leading 
order of QCD using nonperturbative fragmentation functions~[23]. It was shown 
that the transverse momentum and rapidity distributions measured at HERA~[3] 
well agree with theoretical predictions. These comparisons also 
illustrate the significance of the charm component in the resolved photon.
Unfortunately the dijet angular distributions~[2] and correlations~[3]
in charm photoproduction, which give more clear test for the manifestation of 
the relative role of the direct and resoved photon contributions,
were not described in this approach. The NLO QCD predictions in
massive scheme~[20] are in general agreement with data although 
differences have been isolated in regions where contributions from
higher orgers are expected to be significant~[3].

In the present paper we will consider the associated $D^{*}$ and dijet photoproduction 
using the $k_T$-factorization approach. There are several motivations for such a 
study.
 First of all, in the framework of  $k_T$-factorization approach it was 
demonstrated~[24, 25] that resolved photon-like contributions are 
effectively simulated by gluon evolution in the initial state and are described 
by unintegrated gluon distribution in the proton.
 The fraction $x_\gamma^{\rm obs}$ of the photon momentum 
which participates in the dijet production has been measured in Refs.~[1--3].
This quantity is sensitive to the relative contributions of resolved and direct processes 
in collinear fixed-order QCD calculations~[1]. In leading-order (LO), direct photon events
at the parton level have $x_\gamma^{\rm obs} = 1$, while resolved photon
events populate low values of $x_\gamma^{\rm obs}$. The same situation
is observed in a next-to-leading (NLO) calculations, because in the three
parton final state any of these partons are allowed to take any kinematically
accessible value. In the $k_T$-factorization formalism the hardest transverse 
momentum parton emission can be anywhere 
in the evolution chain, and does not need to be closest to the photon as required by the 
strong $\mu^2$ ordering in DGLAP. Thus, if the two hardest jets are produced by the 
$c\bar c$ pair, then $x_\gamma^{\rm obs}$ is close to unity, but if a 
gluon from the initial cascade and one of the final charmed quarks form
the two hardest transverse momentum jets, then $x_\gamma^{\rm obs} < 1$.

Another interesting quantity is the measured distribution of the outgoing jets
with a $D^*$ in the final state on the angle $\theta^*$ between the jet-jet axis 
and the proton beam direction in the dijet rest frame. This quantity
is sensitive to the spin of the propagator in the hard subprocess~[2].
In direct photon processes $\gamma g \to c\bar c$ the propagator 
in the LO QCD diagrams is a quark, and the differential cross section
rises slowly towards high $|\cos \theta^*|$ values. In resolved processes, 
the gluon propagator is allowed at LO and dominates over the quark propagator 
due to the stronger gluon-gluon coupling compared to the quark-gluon coupling. 
If most of the resolved photon charm dijet events are produced as a result of charm 
from the photon, a gluon-exchange contribution in $cg \to cg$ subprocess should 
dominate. This results in a steep rise of the cross section towards high $|\cos \theta^*|$
values. If one of the jets is explicitly tagged as a charm jet, the sign of 
$\cos \theta^*$ can be defined~[2]. In the $k_T$-factorization approach the 
$\cos \theta^*$ distribution is determined only by the photon-gluon fusion off-shell matrix 
element which cover both scattering process. It is because there is no restriction on 
the transverse momenta along the evolution cascade, as it was already mentioned above.

Third, previously unmeasured correlations between the two jets, namely the 
difference in azimuthal angle $\Delta \phi$, and the transverse 
momentum of the dijet system $p_T$, has been presented recently~[3]. 
These quantities are particularly sensitive to high-order correction effects. 
So, in the collinear LO approximation, the two jets are produced
back-to-back with $\Delta \phi = \pi$ and $p_T = 0$.
Large deviations from these values may come from higher-order QCD effects.
Taking into account the non-vanishing initial parton transverse momenta
leads to the violation of this back-to-back kinematics in the 
$k_T$-factorization approach even at leading order. It was shown~[26]
that theoretical and experimental studying of the $\Delta \phi$ distributions is
a direct probe of the non-collinear parton evolution.

In the previous studies~[24, 25] the $k_T$-factorization approach was 
already applied to the calculation of the $x_\gamma^{\rm obs}$
and $|\cos \theta^*|$ distributions of the charm and dijet 
photoproduction at HERA. It was observed the steep rise in 
the cross section with increasing $|\cos \theta^*|$ for resolved photon-like events 
compared to the direct photon-like events through the initial state gluon cascade~[25]. 
It was claimed that this effect in the $k_T$-factorization approach can be interpreted 
as "charm excitation" processes. However, the comparisons with the experimental data 
on $|\cos \theta^*|$ distributions were done~[27] only in the framework of Monte-Carlo 
generator CASCADE~[28]. The azimuthal correlations between the transverse
momenta of the produced jets and the $p_T$ distributions
have not been investigated up to this time.

In this paper we  study the associated $D^*$ and dijet
photoproduction at HERA in more detail. We calculate a number 
of different production rates and compare our theoretical results
with the recent ZEUS data~[1--3]. Special attention will be drawn to
the specific dijet correlations which is sensitive to the transverse momentum 
of the partons incoming to the hard scattering process and therefore sensitive 
to the details of the unintegrated gluon density, as it was mentioned above. 
We will test the unintegrated gluon distributions which are obtained from the 
full CCFM, unified BFKL-DGLAP evolution equations and from the conventional 
(DGLAP-based) gluon densities. In last case we use the KMR prescription.

The outline of  our paper is following. In Section~2 we 
recall the basic formulas of the $k_T$-factorization formalism with a brief 
review of calculation steps. In Section~3 we present the numerical results
of our calculations and a dicussion. Finally, in Section~4, we give
some conclusions.

\section{Basic formulas} \indent 

Let $p_e$ and $p_p$ be the four-momenta of the initial electron and proton,
and $p_c$ and $p_{\bar c}$ the four-momenta of the produced charmed quarks.
The charm photoproduction cross section in the $k_T$-factorization approach
can be written as
$$
  d\sigma(\gamma p \to c\bar c + X) = \int {dx\over x}
 {\cal A}(x,{\mathbf k}_{T}^2,\mu^2) d{\mathbf k}_{T}^2 {d\phi\over 2\pi}
 d{\hat \sigma} (\gamma g^* \to c\bar c), \eqno(1)
$$

\noindent 
where ${\cal A}(x,{\mathbf k}_{T}^2,\mu^2)$ is the unintegrated gluon distribution,
 ${\hat \sigma} (\gamma g^* \to c\bar c)$ is the charm 
production
cross section via an off-shell gluon having fraction $x$ of a initial 
proton longitudinal momentum, non-zero transverse momentum ${\mathbf k}_{T}$ 
(${\mathbf k}_{T}^2 = - k_{T}^2 \neq 0$) and an azimuthal angle $\phi$.
The expression (1) can be easily rewritten as
$$
  { d\sigma (\gamma p \to c\bar c + X) \over dy_c\, d{\mathbf p}_{c\, T}^2 } =
 \int {1\over 16\pi (x s)^2 (1 - \alpha)} {\cal A}(x,{\mathbf k}_T^2,\mu^2) 
  |\bar {\cal M}|^2(\gamma g^* \to c\bar c) d{\mathbf k}_T^2 {d\phi \over 2\pi} {d\phi_c \over 2\pi}, \eqno (2)
$$
\noindent
where $|\bar {\cal M}|^2(\gamma g^* \to c\bar c)$ is squared the off-shell matrix element, 
$s = (p_e + p_p)^2$ is the total center-of-mass frame energy, $y_c$ and $\phi_c$ are 
the rapidity and the azimuthal angle of the produced charmed quark having mass $m_c$, 
$\alpha = m_{c\,T}\exp(y_c)/\sqrt s$ and $m_{c\,T}^2 = m_c^2 + {\mathbf p}_{c\,T}^2$.
The analytic expression for the $|\bar {\cal M}|^2 (\gamma g^* \to c\bar c)$
was obtained in our previous paper~[29]. We would like to note 
that if we average (2) over ${\mathbf k}_{T}$ and take the limit 
${\mathbf k}_{T}^2 \to 0$, then we obtain usual formula for the 
charm production in LO perturbative QCD.

The available experimental data~[1--3] taken by the ZEUS collaboration refer to the 
charm photoproduction in $ep$ collisions, where electron is scattered
at small angle and the mediating photon is almost real ($Q^2 \sim 0$). Therefore
$\gamma p$ cross section (2) needs to be weighted with the photon flux in the electron:
$$
  d\sigma(ep \to c\bar c + X) = \int f_{\gamma/e}(y)dy\, d\sigma(\gamma p \to c\bar c + X), \eqno (3)
$$
\noindent
where $y$ is a fraction of the initial electron energy taken by the photon in the 
laboratory frame, and we use the Weizacker-Williams approximation for the 
bremsstrahlung photon distribution from an electron:
$$
  f_{\gamma/e}(y) = {\alpha_{em} \over 2\pi}\left({1 + (1 - y)^2\over y}\ln{Q^2_{\rm max}\over Q^2_{\rm min}} + 
  2m_e^2 y\left({1\over Q^2_{\rm max}} - {1\over Q^2_{\rm min}} \right)\right). \eqno (4)
$$

\noindent
Here $\alpha_{em}$ is Sommerfeld's fine structure constant, $m_e$ is the electron 
mass, $Q^2_{\rm min} = m_e^2y^2/(1 - y)^2$ and $Q^2_{\rm max} = 1\,{\rm GeV}^2$, 
which is a typical value for the recent photoproduction measurements at the HERA 
collider.

The basic photon-gluon fusion subprocess under consideration ($\gamma g^* \to c\bar c$)
gives rise to two high-energy charmed quarks, which can further evolve into hadron jets.
In our calculations the produced quarks (with their known kinematical 
parameters) were taken to play the role of the final jets. 
These two quarks are accompanied by a number of gluons radiated 
in the course of the gluon evolution. As it has been noted in~[24], on
the average the gluon transverse momentum decreases from the hard interaction
block towards the proton. As an approximation, we assume that the gluon 
emitted in the last evolution step and having the four-momenta $k'$ 
compensates the whole transverse momentum of the gluon participating in the hard 
subprocess, i.e. ${\mathbf k'}_{T} \simeq - {\mathbf k}_{T}$. All the other emitted gluons are 
collected together in the proton remnant, which is assumed to carry only a negligible 
transverse momentum compared to ${\mathbf k'}_{T}$. This gluon gives rise to a 
final hadron jet with $E_T^{\rm jet} = |{\mathbf k'}_{T}|$ in addition to the jet 
produced in the hard subprocess. From these three hadron jets we choose the two ones 
carrying the largest transverse energies, and then compute the charm and 
associated dijet production rates.

As it was noted already, the variable $x_\gamma^{\rm obs}$ is often used
in the analysis of the recent experimental data. This variable, which is the 
fraction of the photon momentum contributing to the production of two jets 
with highest transverse energies $E_T^{\rm jet}$, experimentally is defined as
$$
  x_\gamma^{\rm obs} = { E_T^{{\rm jet}_1} e^{-\eta^{{\rm jet}_1}} + E_T^{{\rm jet}_2} e^{-\eta^{{\rm jet}_2}} \over 2 y E_e}, \eqno (5)
$$

\noindent
where $y E_e$ is the initial photon energy and $\eta^{{\rm jet}_i}$ are
the pseudo-rapidities of these hardest jets. 
The pseudo-rapidities $\eta^{{\rm jet}_i}$ are defined as
$\eta^{{\rm jet}_i} = - \ln \tan (\theta^{{\rm jet}_i}/2)$, where 
$\theta^{{\rm jet}_i}$ are the polar angles of the jets with respect to the proton beam.
The selection of $x_\gamma^{\rm obs} > 0.75$ and $x_\gamma^{\rm obs} < 0.75$ 
yields samples enriched in direct and resolved photon processes, respectively.
The complementary variable is
$$
  x_p^{\rm obs} = { E_T^{{\rm jet}_1} e^{\eta^{{\rm jet}_1}} + E_T^{{\rm jet}_2} e^{\eta^{{\rm jet}_2}} \over 2 E_p}, \eqno (6)
$$

\noindent
which is the fraction of the proton's momentum contributing to the 
production of the two jets. Other dijet variables such as their scattering 
angle $\theta^*$ and invariant mass $M$ are defined as
$$
  \cos \theta^* = \tanh\left( {\eta^{{\rm jet}_1} - \eta^{{\rm jet}_2} \over 2} \right), \eqno (7)
$$
$$
  M = \sqrt { 2 E_T^{{\rm jet}_1} E_T^{{\rm jet}_2} \left[ \cosh(\eta^{{\rm jet}_1} - \eta^{{\rm jet}_2}) - \cos(\phi^{{\rm jet}_1} - \phi^{{\rm jet}_2}) \right] }, \eqno (8)
$$

\noindent
where $\phi^{{\rm jet}_i}$ are the azimuthal angles of the corresponding jets.

The multidimensional integration in (2) and (3) has been performed
by means of the Monte Carlo technique, using the routine 
\textsc{VEGAS}~[30]. The full C$++$ code is available from the authors on 
request\footnote{lipatov@theory.sinp.msu.ru}.

\section{Numerical results} \indent 

We now are in a position to present our numerical results. First we describe our
theoretical input and the kinematical conditions. There are several parameters 
which determined the normalization factor of the cross section under 
consideration: the charmed quark mass $m_c$, factorization and 
normalisation scales $\mu_F$ and $\mu_R$, charm fragmentation function
and unintegrated gluon distribution in a proton ${\cal A}(x,{\mathbf k}_T^2,\mu^2)$. 
In our calculations we convert the charmed quark into a $D^*$ meson using 
the Peterson fragmetation function~[31] with $\epsilon_c = 0.035$~[32]. The branching $c 
\to D^*$ was set to the value measured~[33] by the OPAL collaboration: 
$f(c\to D^*) = 0.235$. 

In the further numerical analysis we have tried three different sets of the 
unintegrated gluon densities in a proton, namely J2003~(set~1)~[34], KMS~[14] 
and KMR~[19], which are frequently discussed in the literature now\footnote{The most 
relevant properties of different unintegrated gluon distributions are 
discussed in~[12, 13].}. The J2003 density has been obtained from the numerical solution 
of the full CCFM equation. The input parameters were fitted~[34] to 
describe the proton structure function $F_2(x,Q^2)$.
The J2003~(set~1) gluon distribution contains only
singular terms in the CCFM splitting function $P_{gg}(z)$\footnote{See Ref.~[34] 
for more details.}. 
This gluon density has been applied, 
in particular, in the analysis of the forward jet production at HERA, charm and 
bottom production at Tevatron~[32], and charm and $J/\psi$ production at 
LEP2 energies~[27]. Another set (KMS) is obtained~[14] from a unified 
BFKL-DGLAP description of $F_2(x, Q^2)$ data and includes the so-called 
consistency constraint~[35]. The consistency constraint introduces a 
large correction to the LO BFKL equation. It was argued~[36] that 
about 70\% of the full next-to-leading (NLO) corrections to the BFKL exponent 
$\Delta$ are effectively included in this constraint. The last unintegrated 
gluon distribution used here (the so-called KMR distribution) is the one which was 
originally proposed in~[19]. The KMR approach is the formalism to construct unintegrated 
gluon distribution from the known conventional parton densities $xa(x,\mu^2)$, where 
$a = g$ or $a = q$. It accounts for the angular-ordering (which comes from the coherence 
effects in gluon emission) as well as the main part of the collinear higher-order QCD 
corrections. The KMR-constructed parton densities were used, in particular, to 
describe the heavy quark and $J/\psi$ meson production in $\gamma \gamma$ collisions 
at LEP2~[29, 37] and prompt photon in photo- and hadroproduction 
at HERA and Tevatron~[38, 39].

The most significant theoretical uncertainties come also from the
choice of the factorization and renormalization scales. First of them
is related to the evolution of the gluon distributions 
${\cal A}(x,{\mathbf k}_T^2,\mu_F^2)$, the other is responsible 
for the strong coupling constant $\alpha_s(\mu^2_R)$.
As it often done~[20] for charm production, we choose the renormalization and 
factorization scales to be equal: $\mu_R = \mu_F = \mu = \sqrt{m_c^2 + \langle {\mathbf p}_{T}^2 \rangle}$, 
where $\langle {\mathbf p}_{T}^2 \rangle$ was set to the average ${\mathbf p}_{T}^2$ 
of the charm quark and antiquark\footnote{We use special choice $\mu^2 = {\mathbf k}_T^2$ 
in the case of KMS gluon, as it was originally proposed in~[14].}. 
In the present paper we concentrate mostly on the non-collinear gluon 
evolution in the proton and do not study the scale dependence of our 
results. 
To completeness, the charm mass was set to 
$m_c = 1.4$~GeV and we use LO formula for the coupling constant $\alpha_s(\mu^2)$ 
with $n_f = 4$ active quark flavours and $\Lambda_{\rm QCD} = 200$~MeV, such 
that $\alpha_s(M_Z^2) = 0.1232$.

The recent experimental data~[1--3] for the associated $D^*$ and dijet photoproduction
at HERA come from ZEUS collaboration. 
The data~[1] refer to the kinematical region\footnote{Here and in the following 
all kinematic quantities are given in the laboratory frame where positive OZ axis 
direction is given by the proton beam.} defined by $130 < W < 280$~ GeV, 
$Q^2 < 1$~GeV$^2$ and given
for jets with $|\eta^{\rm jet}| < 2.4$, $E_T^{{\rm jet}_1} > 7$~GeV, 
$E_T^{{\rm jet}_2} > 6$~GeV and at least one $D^*$ in the range $p_T^{D^*} > 3$~GeV,
$-1.5 < \eta^{D^*} < 1.5$. Results are also presented for the region
$E_T^{{\rm jet}_1} > 6$~GeV, $E_T^{{\rm jet}_2} > 5$~GeV.
The data~[2] have been obtained in the kinematic range
$130 < W < 280$~ GeV, $Q^2 < 1$~GeV$^2$, $p_T^{D^*} > 3$~GeV,
$-1.5 < \eta^{D^*} < 1.5$, $E_T^{\rm jet} > 5$~GeV and $|\eta^{\rm jet}| < 2.4$.
The cuts on the dijet invariant mass $M > 18$~GeV and on the 
average jet pseudorapidity $|\bar \eta^{\rm jet}| < 0.7$ were applied, 
where $\bar \eta^{\rm jet}$ is defined as 
$\bar \eta^{\rm jet} = (\eta^{{\rm jet}_1} + \eta^{{\rm jet}_2})/2$.
The more recent data~[3] refer to the kinematical region defined by
$E_T^{{\rm jet}_1} > 7$~GeV, $E_T^{{\rm jet}_2} > 6$~GeV and 
$-1.5 < \eta^{\rm jet} < 2.4$. The $Q^2$, $W$, $p_T^{D^*}$ and $\eta^{D^*}$
requirements are the same as in the previous measurements.

\subsection{The distributions on  $x_\gamma^{\rm obs}$ and $x_p^{\rm obs}$}
 \indent

In Figs.~1 --- 5 we confront the $x_\gamma^{\rm obs}$ and $x_p^{\rm obs}$ distributions 
calculated in different kinematical regions with the ZEUS data.
The solid, dashed and dash-dotted histograms 
correspond to the results obtained with the J2003~(set~1), KMR and 
KMS unintegrated gluon densities, respectively.
In agreement with the expectation for direct photon processes, the peak 
in $x_\gamma^{\rm obs}$ distributions at high values of the 
$x_\gamma^{\rm obs}$ is observed both in data and in theoretical 
calculations. However, there is also a substantial tail to small values 
of $x_\gamma^{\rm obs}$. As it was mentioned above, the existence of this plateau 
in the collinear approximation of QCD usually is attributted~[1--3] to the charm 
excitation from a resolved photon and is interpreted as a likely signature of the 
photon structure. In the $k_T$-factorization approach this
plateau indicates the fact that gluon radiated from the evolution
cascade appears to be harder than charmed quarks 
(produced in hard parton interaction) in a significant fraction of events.
Since in our calculations we have not included the resolved photon contribution 
explicitly and have operated in terms of the proton structure only, we can 
conclude that the $k_T$-factorization approach effectively imitates the charm 
component of the photon~[24, 25]. However, the predicted tail at 
small $x_\gamma^{\rm obs}$ values is strongly depends on the 
unintegrated gluon distributions. Our results corresponding to 
different gluon densities do not agree well with the ZEUS data 
in Figs.~1,~2 and~4. The calculated cross sections at low
$x_\gamma^{\rm obs}$ are defined by the average value of the gluon
transverse momenta $\langle k_{T} \rangle$ which is generated in the
course of the non-colliner evolution. It is because the 
events when the gluon jet has the largest and next-to-largest $p_T$ among the 
three hadron jets contribute only in this kinematical region~[24].
Therefore we can conclude that average gluon $\langle k_{T} \rangle$ which generated by 
the all three versions of the unintegrated gluon distributions under 
consideration is 
too small to describe the ZEUS data. However, Fig.~3 shows that the theoretical 
results obtained with the J2003 and KMR unintegrated gluon 
distributions well describe the experimental data with the cut on the dijet 
invariant mass $M > 18$~GeV. It demonstrates that this cut is essential for applicability 
of the description of resoved photon contributions by noncollinear evolution only.

Note that collinear NLO massive calculations~[20] give the similar description of the 
ZEUS data for $d\sigma/x_\gamma^{\rm obs}$: the cross sections 
predicted by the NLO calculations reproduce the data in direct photon-like
region but they are below the data in resolved photon-like one [1--3]. 
This fact is clearly demonstrates again that $k_T$-factorization
approach effectively simulates charm quark excitation processes which 
give a main contribution to the NLO cross section at low $x_\gamma^{\rm obs}$. 
Note also that in according to the 
analysis~[2, 3] which was done by the ZEUS collaboration, in order to obtain a 
realistic comparison of their data and theory the corrections for hadronisation 
should be taken into account in the predictions\footnote{See Refs.~[2, 3] for more details.}. 
The correction factors are typically 0.9 -- 1.1 depending on a bin.
These factors are not accounted for in our analysis.

The similar description of the data~[1] was obtained in~[25] where the JS 
unintegrated gluon density~[40] and Monte-Carlo generator \textsc{CASCADE}~[28] 
have been applied. However, our predictions lie significantly below results 
presented in~[24]. The reason of this discrepancy are connected with the
parameter settings accepted in~[24]. In particular, in~[24] the 
unintegrated gluon distribution proposed by Bl\"umlein~[41] has been used.

The differential cross section as a function of $x_p^{\rm obs}$ is shown
in Fig.~5. The shape of $x_p^{\rm obs}$ distribution is well reproduced by all 
unintegrated gluon densities under interest. 
However, the KMS gluon distribution (which is successful in description of the 
bottom production at Tevatron~[26, 42] and deep inelastic $J/\psi$ production at HERA~[43]) 
significantly overestimates the data at low values of $x_p^{\rm obs}$, namely 
$x_p^{\rm obs} < 0.02$. This fact is connected with the special choice 
of the renormalization scale $\mu^2 = {\mathbf k}_{T}^2$ in the running 
coupling constant. The J2003~(set~1) and KMR gluon 
densities are in good agreement with data. 
Note that the measured cross section $d\sigma/dx_p^{\rm obs}$ are also well 
described by the massive NLO QCD predictions. However the data tend to agree 
better with the upper bound of these calculations.

\subsection{Angular distributions} \indent

Figs.~6 and~7 show the differential cross section as a function of
$|\cos \theta^*|$ separately for the direct-enriched ($x_\gamma^{\rm obs} > 0.75$)
and resolved-enriched ($x_\gamma^{\rm obs} < 0.75$) samples.
As it was mentioned above, studying of these distributions give us the possibility
to learn about the size of the contribution from different production mechanisms.
It is because the angular dependence of the subprocess involving a
gluon propagator in the $t$ channel is approximately proportional
to $(1 - |\cos \theta^*|)^{-2}$, whereas it is proportional
to $(1 - |\cos \theta^*|)^{-1}$ in the case of quark propagator.
So, from Fig.~6 one can see that the direct photon-like events give a slow increase 
in cross section with increasing $|\cos \theta^*|$ (in proton direction) 
both in the data and 
in the theoretical calculations. The resolved photon-like events 
exhibids a more rapid rise towards high values of $|\cos \theta^*|$ (Fig.~7).
Such a behaviour is suggested from the large 
gluon-exchange contribution of charm-excitation process. 
In our theoretical calculations, the shape of the data is reproduced very well
but overall normalisation is rather low compared to the data.
Note that the collinear NLO predictions~[20] are also significant below 
the data at low $x_\gamma^{\rm obs}$~[2].

In the further analysis~[2] which was done by
the ZEUS collaboration, the two jets were distinguished by
associating the $D^*$ meson to the closest jet in $\eta$ --- $\phi$ plane.
The associated jet is defined as the jet with the smallest
$R_i^2 = (\eta^{{\rm jet}_i} - \eta^{D^*})^2 + (\phi^{{\rm jet}_i} - \phi^{D^*})^2$,
where $\phi^{{\rm jet}_i}$ and $\phi^{D^*}$ are the azimuthal angles
of the jets and the $D^*$ meson in the laboratory frame.
Calling this "associated jet" jet~1 in~(7), the rise of $d\sigma/d\cos \theta^*$
can be studied~[2]. Figs.~8 and~9 show the differential cross section as a 
function of $\cos \theta^*$ for the direct-enriched and resolved-enriched samples.
The resolved photon-like events exhibit a mild rise in the proton hemisphere 
towards $\cos\theta^* = 1$, consistent with expectations from quark exchange.
In contrast, they have a strong rise towards $\cos\theta^* = - 1$, i.e. in
photon direction, consistent with a dominant contribution from
gluon exchange. In our theoretical calculations, the peak at 
$\cos\theta^* = - 1$ at low $x_\gamma^{\rm obs}$ clearly illustrates
again that the $k_T$-factorization approach effectively reproduces the 
charm excitation processes using only the photon-gluon fusion off-mass shell 
matrix elements. It is necessary to note that these matrix elements 
correspond to the $2 \to 2$ partonic subprocess with the charm-anticharm pair 
in final state and therefore is fully symmetric in $\cos\theta^*$. 
This fact leads to the symmetric $\cos\theta^*$ distribution
at high $x_\gamma^{\rm obs}$ (Fig.~8). 
However, the angular distribution $d\sigma/d\cos \theta^*$ exhibits a 
slight asymmetry in the data as well in the NLO predictions which 
is explained~[2] by the feedthrough from the resolved photon processes near 
$\cos\theta^* = - 1$. So we can conclude that all three
unintegrated gluon densities studied here overestimate the data at
high values of $\cos\theta^*$ and $x_\gamma^{\rm obs}$.

\subsection{The invariant mass distributions and azimuthal correlations} \indent

Very recently the ZEUS collaboration has been measured [3]  
the cross sections of the $D^*$ meson and associated dijet production 
as function of dijet invariant mass $M$, and the correlations between final 
hadronic jets, namely the difference in azimuthal angle 
$\Delta \phi = |\phi^{{\rm jet}_1} - \phi^{{\rm jet}_2}|$, 
and the transverse momentum $p_T$ distributions of the dijet system 
(${\mathbf p}_{T} = {\mathbf p}_{T}^{{\rm jet}_1} + {\mathbf p}_{T}^{{\rm jet}_2}$). 
As it was mentioned above, the $\Delta \phi$ and $p_T$ distributions 
are particularly sensitive to high-order corrections and to unintegrated gluon 
densities in the proton. In Figs.~10 --- 18 the differential dijet cross
sections as a function of these variables are shown in different 
$x_\gamma^{\rm obs}$ regions. We see again that the agreement between the theoretical 
calculations and the data is better for the direct-enriched events in comparison
with resolved-enriched ones. In spite of the fact all 
histograms in Figs.~13 --- 15 lie above the data at $\Delta \phi \sim 0$, 
the shape of $\Delta \phi$ distribution at $x_\gamma^{\rm obs} > 0.75$ 
strongly depends on the unintegrated gluon densities used.
The J2003 density gives a significantly harder distribution compared to the 
data whereas the KMS one gives more softer distribution.
In contrast, in the low $x_\gamma^{\rm obs}$ region the shapes of $\Delta \phi$ 
spectrum predicted by the different gluon distributions 
are very similar to each other. Therefore by analogy with the bottom production
at Tevatron~[26] we can conclude that the properties of different
unintegrated gluon densities manifest themselves in the 
dijet azimuthal correlations at high values of $x_\gamma^{\rm obs}$.
Concerning the $p_T$-spectra, we see in Figs.~16 --- 18 that our predictions 
have a significantly softer $p_T$ distribution at large $p_T$ compared to the 
data for both direct and resolved photon events. In fact, a reasonable agreement
with ZEUS data in the restricted $p_T$ region ($p_T < 10$~GeV) can be obtained 
using the J2003~(set~1) gluon only. The shape of this distribution 
is also very different for different unintegrated gluon densities. 
The KMS gluon distribution gives the very soft $p_T$ spectrum
in comparison with the J2003 and KMR densities and significantly (by a factor 
about 3) overestimate the data at low $p_T$ (except small $x_\gamma^{\rm obs}$ 
region). The collinear NLO predictions~[20] at high $x_\gamma^{\rm obs}$ 
also show a large deviation from the measured cross sections $d\sigma/d\Delta \phi$ and 
$d\sigma/d p_T$ at low $\Delta \phi$ and high $p_T$~[3]. This discrepancy is
essentially enhanced to the resolved-enriched events. Since the small 
$x_\gamma^{\rm obs}$ region is expected to be particularly sensitive to high-order 
corrections, the further theoretical attempts to reduce observed discrepancy are 
necessary.

Finally, we can conclude that results presented here clearly demonstrate that 
agreement between the theoretical calculations and recent ZEUS data 
for charm production at HERA is far from ideal and for many observables coincide
with the NLO results. We have obtained a rather well description of the 
HERA data with 
the J2003~(set~1) and KMR unintegrated gluon 
distributions in direct photon-like region, but faulty description for many 
observables in resolved photon-like photon region. In the framework of the 
$k_T$-factorization, the different unintegrated gluon densities exhibit to 
significantly different effects at HERA energies. This facts indicates the need for 
better experimental constraints as well as further theoretical studies for a more 
detailed understanding of parton evolution at high energies and, in particular, 
for the precise description of charm with associated jets 
photoproduction at HERA.

\section{Conclusions} \indent 

We presented the calculations of the charm and dijet associated photoproduction
at HERA energies in the $k_T$-factorization approach. We used
the unintegrated gluon densities in a proton which are obtained from
the full CCFM, from unified BFKL-DGLAP evolution equations as well as 
from the Kimber-Martin-Ryskin prescription. The ability of these
$k_T$-dependent gluon densities to reproduce the recent experimental
data taken by the ZEUS collaboration has been investigated.
The calculations of the number of dijet correlations in the 
framework of the $k_T$-factorization were performed for the first time. 

Our investigations were based on the leading-order off-mass shell 
matrix elements for the photon-gluon fusion suprocess.
We have shown that these matrix elements combined with
the non-collinear evolution of gluon densities in a proton
effectively simulate the charmed quark excitation processes and 
indeed the hardest $p_T$ emission comes frequently
from a gluon in the initial-state gluon cascade\footnote{In this part
our conclusions coincide with the ones from~[24, 25].}.
We demostrated that wide plateau seen in $x_\gamma^{\rm obs}$ distributions 
(usually attributed to the charm excitation from a resolved photon)
is connected with the average value of the gluon transverse momenta 
$\langle k_{T} \rangle$ which generates in the course of the 
non-colliner evolution. Special attention has been drawn to the specific 
angular correlations between the hadronic jets in final state. We find that 
the properties of different unintegrated gluon densities manifest themselves 
in the dijet azimuthal correlations at high $x_\gamma^{\rm obs}$.

The absolute cross sections predicted by the $k_T$-factorization calculation
supplemented with the J2003~(set~1) and KMR unintegrated gluon 
distributions reproduce the numerous HERA data for the sample enriched in direct 
photons but are below for the sample enriched in resoved photons.
Therefore further theoretical studies for more detailed understanding of parton 
evolution in a proton in the small-$x$ region are necessary in order to 
describe the charm with associated dijet photoproduction at HERA.

\section{Acknowledgements} \indent 

The authors are very grateful to S.P.~Baranov for encouraging interest
and  helpful discussions, H. Jung for reading of the manuscript and
very useful remarks, P.F. Ermolov for support and DESY Directorate for 
hospitality and support. This research was supported in part by the 
FASI of Russian Federation (grant NS-1685.2003.2).

\begin{figure}
\epsfig{figure=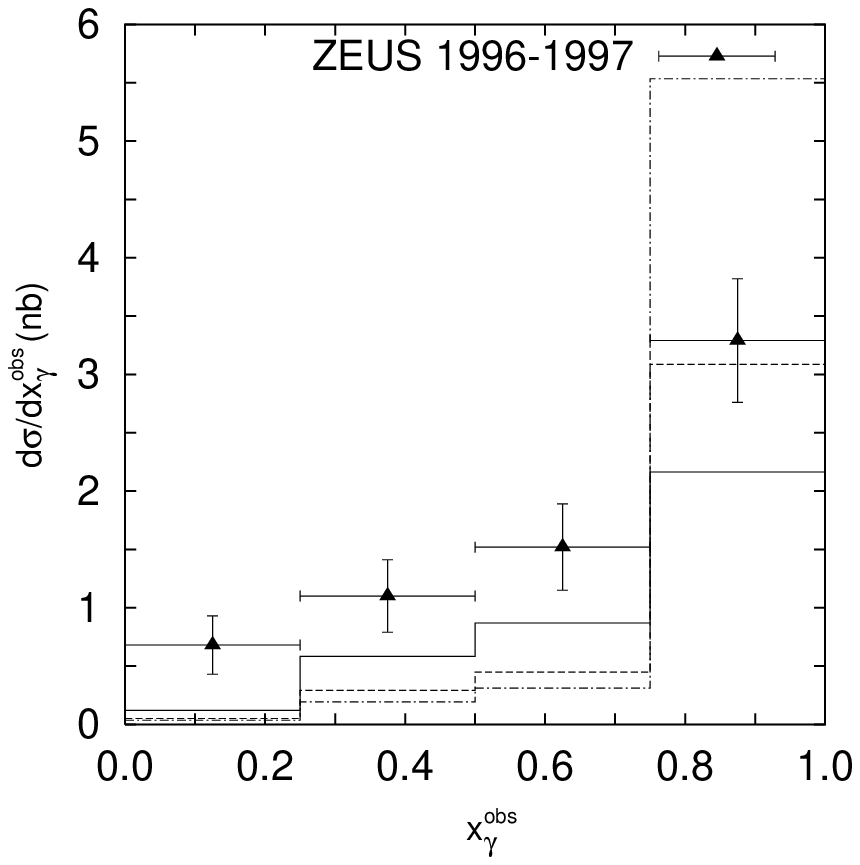, width = 22cm}
\caption{The differential cross section $d\sigma/d x_\gamma^{\rm obs}$
for dijets with an associated $D^*$ meson with
$p_T^{D^*} > 3$~GeV, $-1.5 < \eta^{D^*} < 1.5$ in the
kinematic range $130 < W < 280$~GeV, $Q^2 < 1$~GeV$^2$, 
$|\eta^{\rm jet}| < 2.4$, $E_T^{{\rm jet}_1} > 7$~GeV 
and $E_T^{{\rm jet}_2} > 6$~GeV. The solid, dashed and dash-dotted
histograms correspond to the J2003~(set~1), KMR and KMS unintegrated gluon 
distributions, 
respectively. The experimental data are from ZEUS~[1].}
\label{fig1}
\end{figure}

\newpage

\begin{figure}
\epsfig{figure=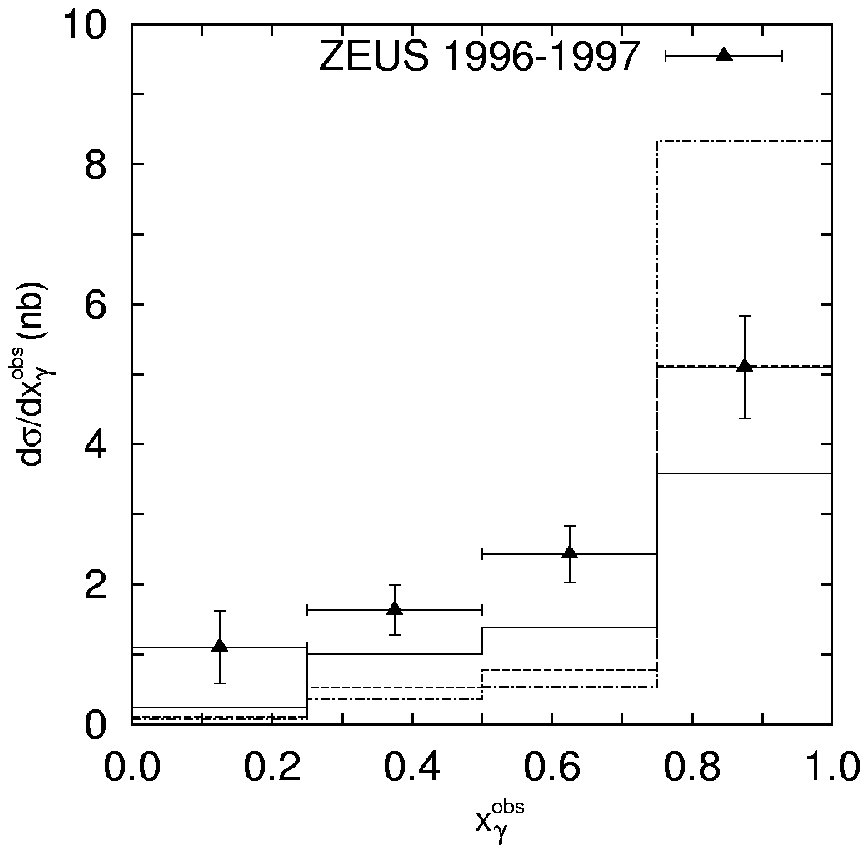, width = 22cm}
\caption{The differential cross section $d\sigma/d x_\gamma^{\rm obs}$
for dijets with an associated $D^*$ meson with
$p_T^{D^*} > 3$~GeV, $-1.5 < \eta^{D^*} < 1.5$ in the
kinematic range $130 < W < 280$~GeV, $Q^2 < 1$~GeV$^2$, 
$|\eta^{\rm jet}| < 2.4$, $E_T^{{\rm jet}_1} > 6$~GeV 
and $E_T^{{\rm jet}_2} > 5$~GeV. The histograms
are the same as in Fig.~1. The experimental data are from ZEUS~[1].}
\label{fig2}
\end{figure}

\newpage

\begin{figure}
\epsfig{figure=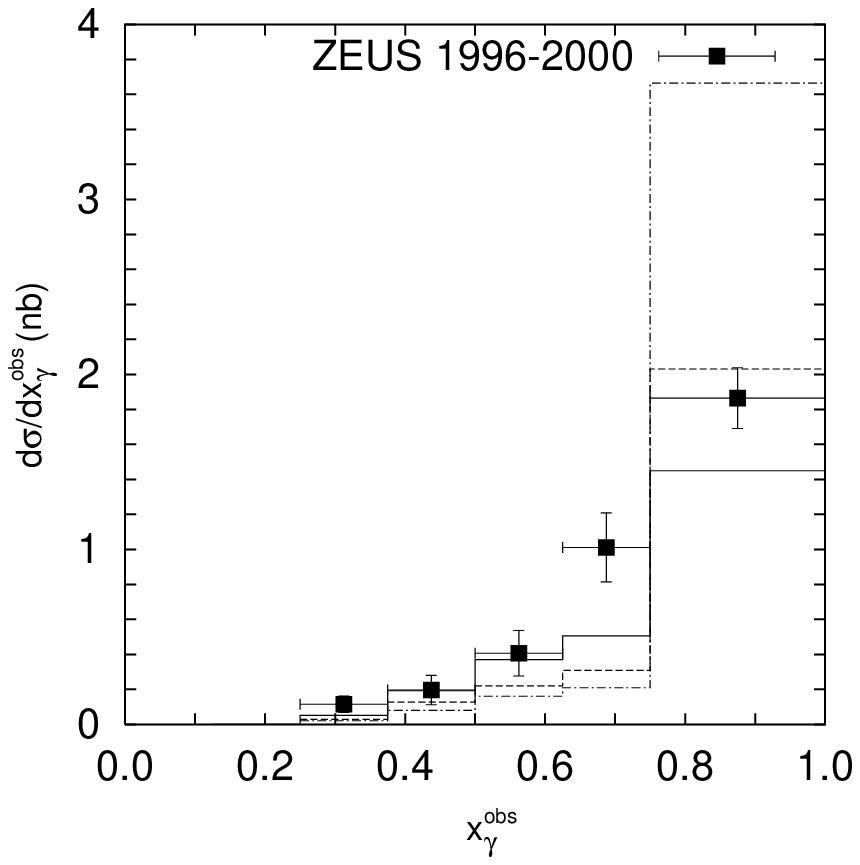, width = 22cm}
\caption{The differential cross section $d\sigma/d x_\gamma^{\rm obs}$
for dijets with an associated $D^*$ meson with
$p_T^{D^*} > 3$~GeV, $-1.5 < \eta^{D^*} < 1.5$ in the
kinematic range $130 < W < 280$~GeV, $Q^2 < 1$~GeV$^2$, 
$|\eta^{\rm jet}| < 2.4$, $E_T^{\rm jet} > 5$~GeV, $M > 18$~GeV and
$|\bar \eta^{\rm jet}| < 0.7$. The histograms
are the same as in Fig.~1. The experimental data are from ZEUS~[2].}
\label{fig3}
\end{figure}

\newpage

\begin{figure}
\epsfig{figure=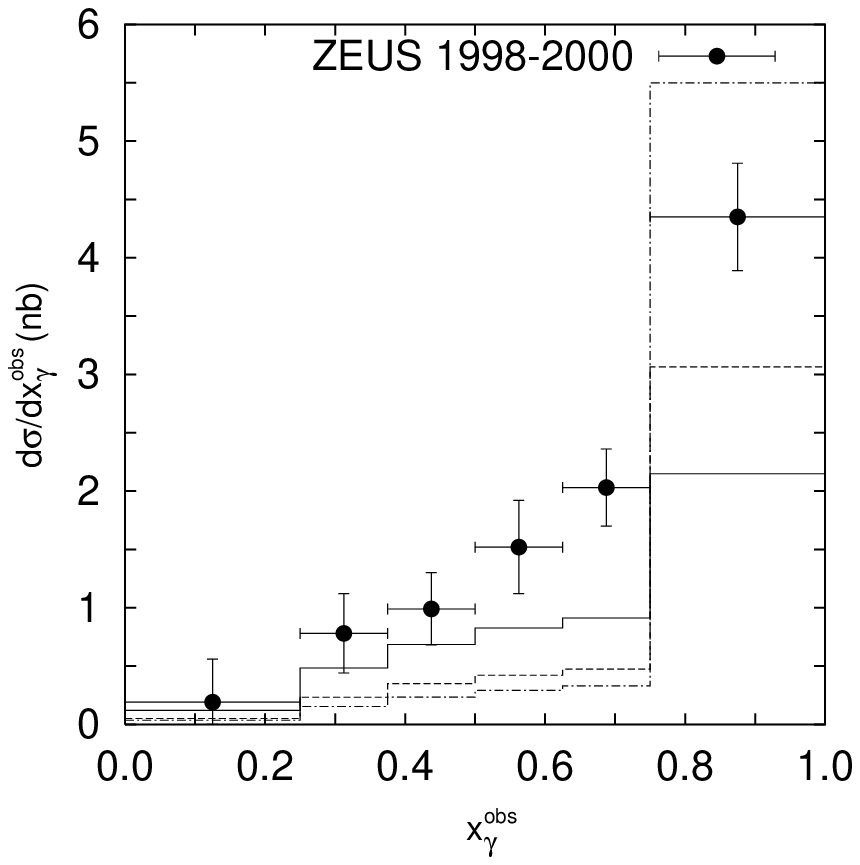, width = 22cm}
\caption{The differential cross section $d\sigma/d x_\gamma^{\rm obs}$
for dijets with an associated $D^*$ meson with
$p_T^{D^*} > 3$~GeV, $-1.5 < \eta^{D^*} < 1.5$ in the
kinematic range $130 < W < 280$~GeV, $Q^2 < 1$~GeV$^2$, 
$-1.5 < \eta^{\rm jet} < 2.4$, $E_T^{{\rm jet}_1} > 7$~GeV 
and $E_T^{{\rm jet}_2} > 6$~GeV. The histograms
are the same as in Fig.~1. The experimental data are from ZEUS~[3].}
\label{fig4}
\end{figure}

\newpage

\begin{figure}
\epsfig{figure=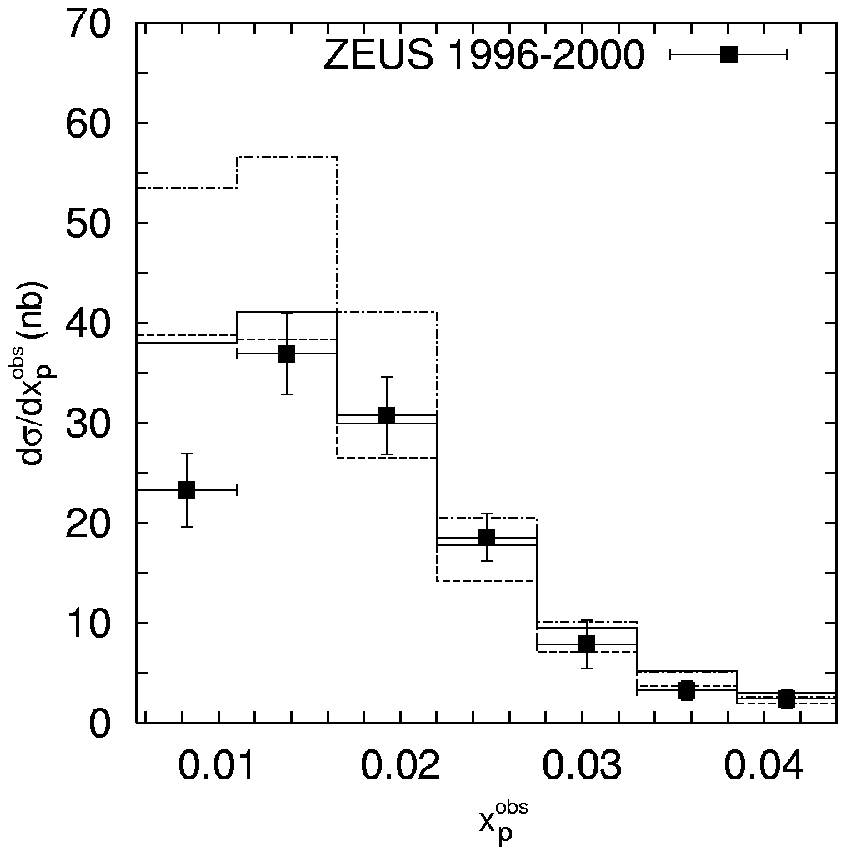, width = 22cm}
\caption{The differential cross section $d\sigma/d x_p^{\rm obs}$
for dijets with an associated $D^*$ meson with
$p_T^{D^*} > 3$~GeV, $-1.5 < \eta^{D^*} < 1.5$ in the
kinematic range $130 < W < 280$~GeV, $Q^2 < 1$~GeV$^2$, 
$|\eta^{\rm jet}| < 2.4$, $E_T^{\rm jet} > 5$~GeV, $M > 18$~GeV and
$|\bar \eta^{\rm jet}| < 0.7$. The histograms
are the same as in Fig.~1. The experimental data are from ZEUS~[2].}
\label{fig5}
\end{figure}

\newpage

\begin{figure}
\epsfig{figure=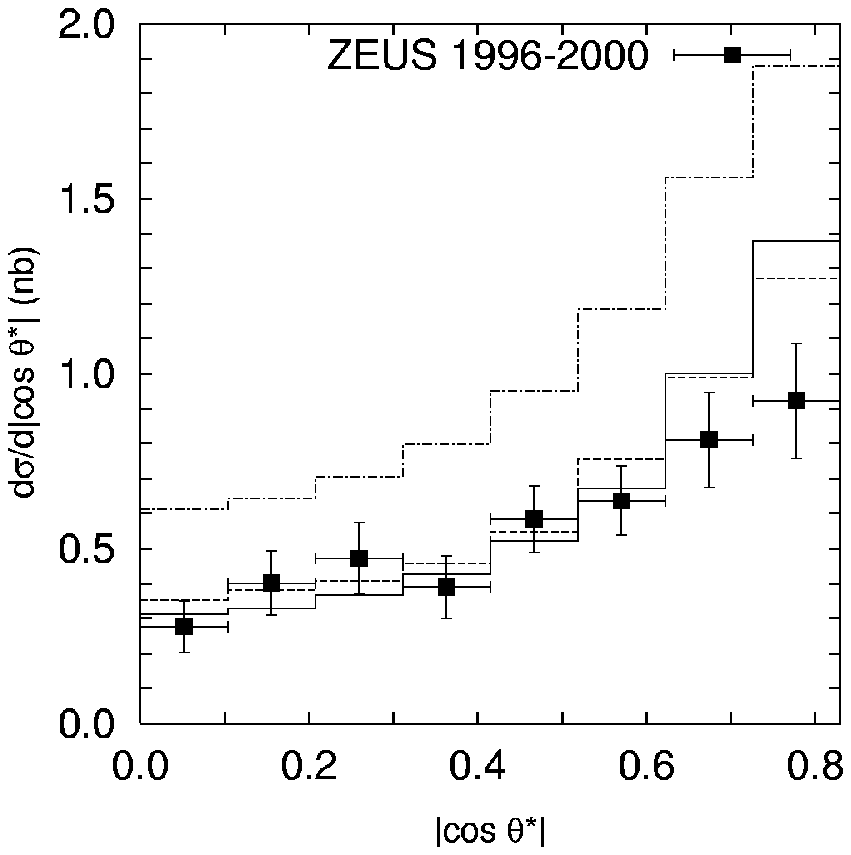, width = 22cm}
\caption{The differential cross section $d\sigma/d|\cos\theta^*|$
for dijets with an associated $D^*$ meson with
$p_T^{D^*} > 3$~GeV, $-1.5 < \eta^{D^*} < 1.5$ in the
kinematic range $130 < W < 280$~GeV, $Q^2 < 1$~GeV$^2$, 
$|\eta^{\rm jet}| < 2.4$, $E_T^{\rm jet} > 5$~GeV, $M > 18$~GeV,
$|\bar \eta^{\rm jet}| < 0.7$ and $x_\gamma^{\rm obs} > 0.75$. The 
hostograms
are the same as in Fig.~1. The experimental data are from ZEUS~[2].}
\label{fig6}
\end{figure}

\newpage

\begin{figure}
\epsfig{figure=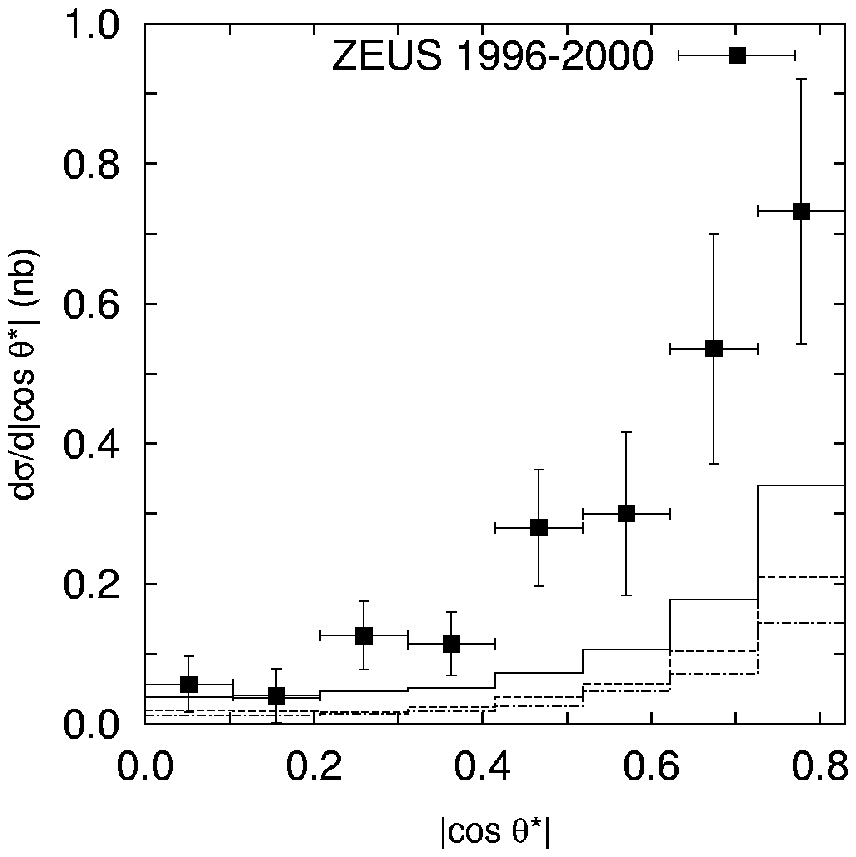, width = 22cm}
\caption{The differential cross section $d\sigma/d|\cos\theta^*|$
for dijets with an associated $D^*$ meson with
$p_T^{D^*} > 3$~GeV, $-1.5 < \eta^{D^*} < 1.5$ in the
kinematic range $130 < W < 280$~GeV, $Q^2 < 1$~GeV$^2$, 
$|\eta^{\rm jet}| < 2.4$, $E_T^{\rm jet} > 5$~GeV, $M > 18$~GeV, 
$|\bar \eta^{\rm jet}| < 0.7$ and $x_\gamma^{\rm obs} < 0.75$. The 
histograms
are the same as in Fig.~1. The experimental data are from ZEUS~[2].}
\label{fig7}
\end{figure}

\newpage

\begin{figure}
\epsfig{figure=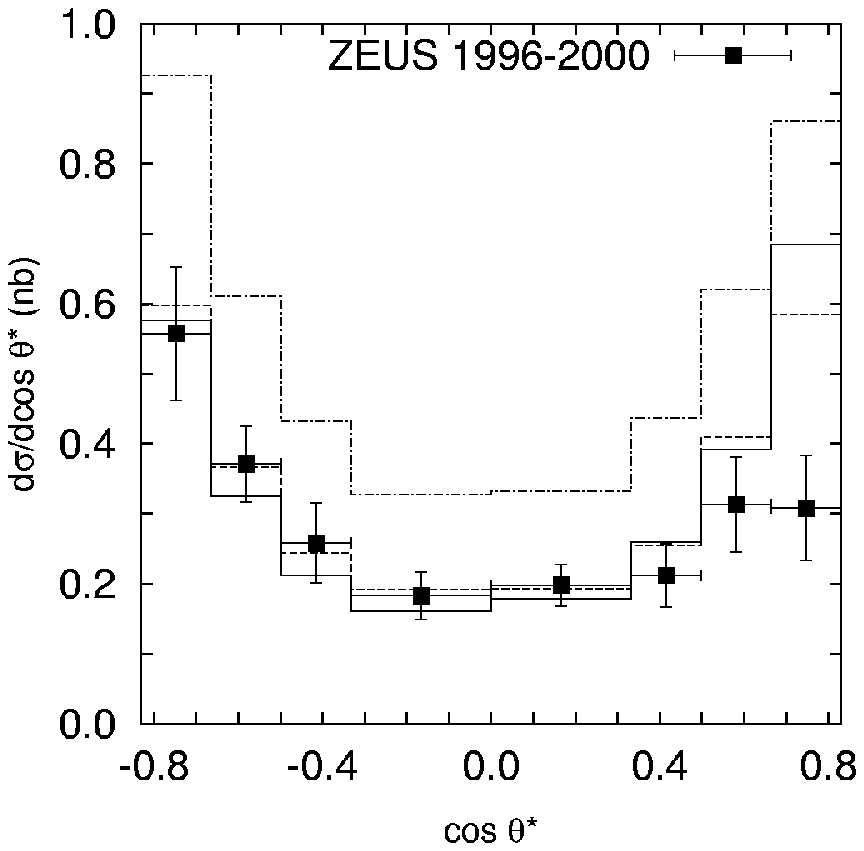, width = 22cm}
\caption{The differential cross section $d\sigma/d\cos\theta^*$
for dijets with an associated $D^*$ meson with
$p_T^{D^*} > 3$~GeV, $-1.5 < \eta^{D^*} < 1.5$ in the
kinematic range $130 < W < 280$~GeV, $Q^2 < 1$~GeV$^2$, 
$|\eta^{\rm jet}| < 2.4$, $E_T^{\rm jet} > 5$~GeV, $M > 18$~GeV, 
$|\bar \eta^{\rm jet}| < 0.7$ and $x_\gamma^{\rm obs} > 0.75$. The 
histograms
are the same as in Fig.~1. The experimental data are from ZEUS~[2].}
\label{fig8}
\end{figure}

\newpage

\begin{figure}
\epsfig{figure=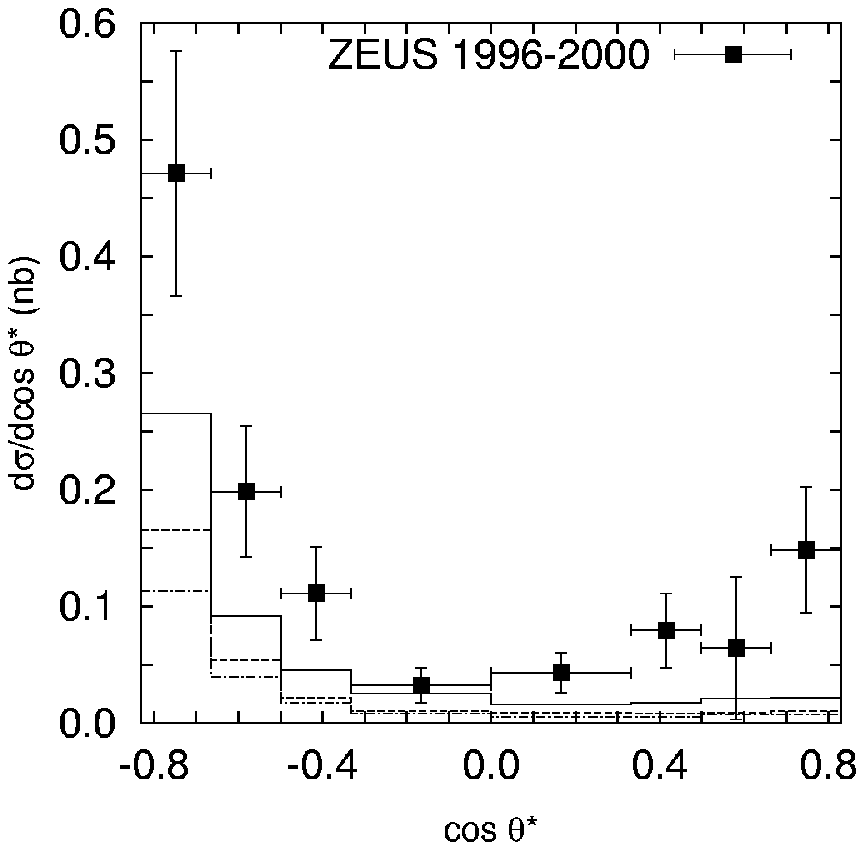, width = 22cm}
\caption{The differential cross section $d\sigma/d\cos\theta^*$
for dijets with an associated $D^*$ meson with
$p_T^{D^*} > 3$~GeV, $-1.5 < \eta^{D^*} < 1.5$ in the
kinematic range $130 < W < 280$~GeV, $Q^2 < 1$~GeV$^2$, 
$|\eta^{\rm jet}| < 2.4$, $E_T^{\rm jet} > 5$~GeV, $M > 18$~GeV, 
$|\bar \eta^{\rm jet}| < 0.7$ and $x_\gamma^{\rm obs} < 0.75$. The 
histograms
are the same as in Fig.~1. The experimental data are from ZEUS~[2].}
\label{fig9}
\end{figure}

\newpage

\begin{figure}
\epsfig{figure=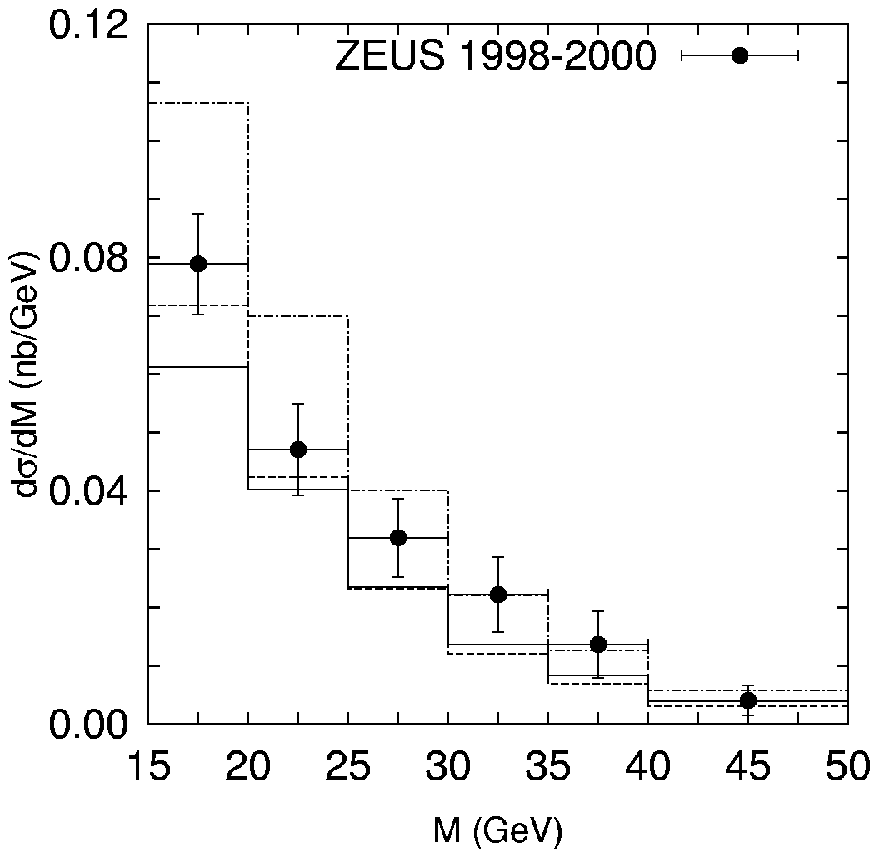, width = 22cm}
\caption{The differential cross section $d\sigma/d M$
for dijets with an associated $D^*$ meson with
$p_T^{D^*} > 3$~GeV, $-1.5 < \eta^{D^*} < 1.5$ in the
kinematic range $130 < W < 280$~GeV, $Q^2 < 1$~GeV$^2$, 
$-1.5 < \eta^{\rm jet} < 2.4$, $E_T^{{\rm jet}_1} > 7$~GeV, 
$E_T^{{\rm jet}_2} > 6$~GeV and $x_\gamma^{\rm obs} > 0.75$. The histograms
are the same as in Fig.~1. The experimental data are from ZEUS~[3].}
\label{fig10}
\end{figure}

\newpage

\begin{figure}
\epsfig{figure=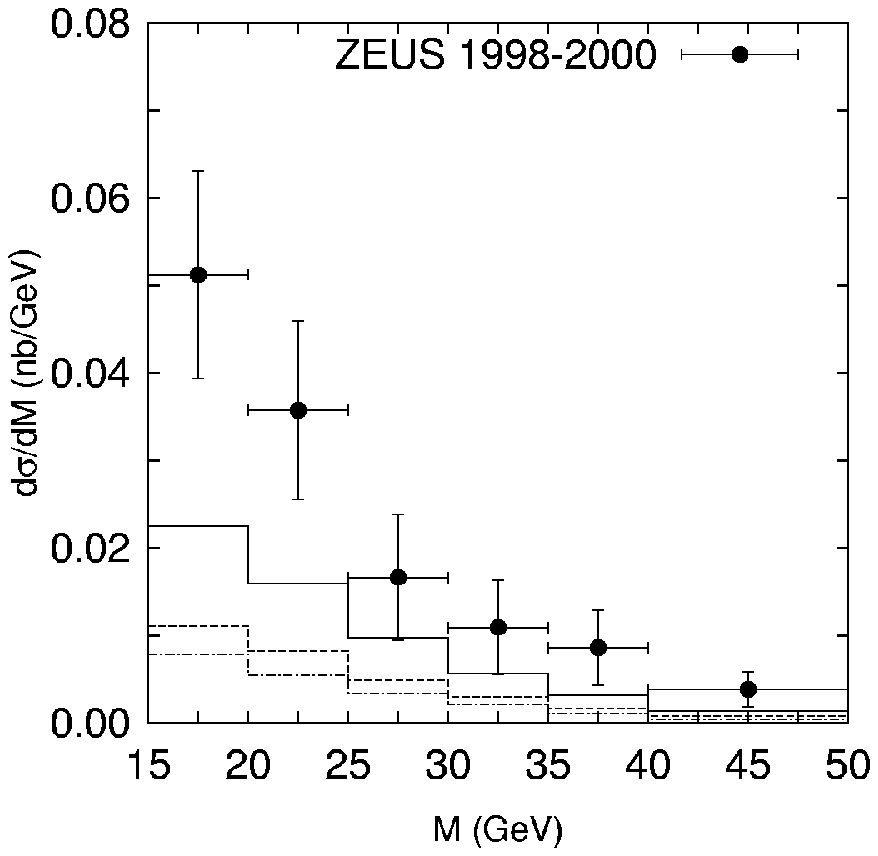, width = 22cm}
\caption{The differential cross section $d\sigma/d M$
for dijets with an associated $D^*$ meson with
$p_T^{D^*} > 3$~GeV, $-1.5 < \eta^{D^*} < 1.5$ in the
kinematic range $130 < W < 280$~GeV, $Q^2 < 1$~GeV$^2$, 
$-1.5 < \eta^{\rm jet} < 2.4$, $E_T^{{\rm jet}_1} > 7$~GeV, 
$E_T^{{\rm jet}_2} > 6$~GeV and $x_\gamma^{\rm obs} < 0.75$. The histograms
are the same as in Fig.~1. The experimental data are from ZEUS~[3].}
\label{fig11}
\end{figure}

\newpage

\begin{figure}
\epsfig{figure=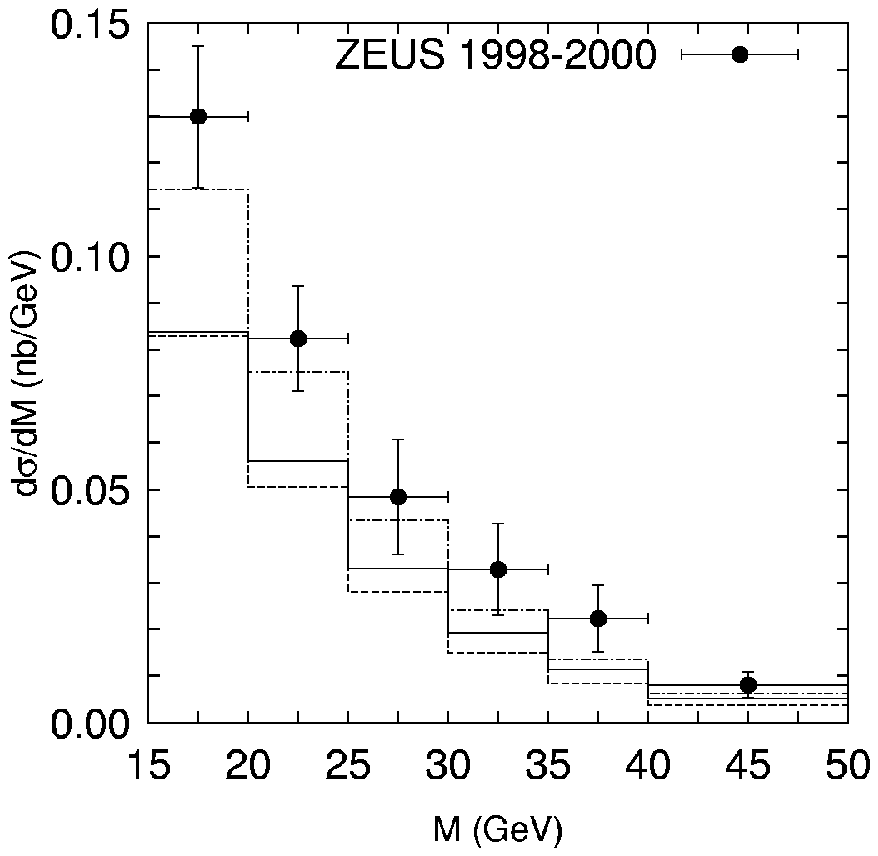, width = 22cm}
\caption{The differential cross section $d\sigma/d M$
for dijets with an associated $D^*$ meson with
$p_T^{D^*} > 3$~GeV, $-1.5 < \eta^{D^*} < 1.5$ in the
kinematic range $130 < W < 280$~GeV, $Q^2 < 1$~GeV$^2$, 
$-1.5 < \eta^{\rm jet} < 2.4$, $E_T^{{\rm jet}_1} > 7$~GeV 
and $E_T^{{\rm jet}_2} > 6$~GeV. The histograms
are the same as in Fig.~1. The experimental data are from ZEUS~[3].}
\label{fig12}
\end{figure}

\newpage

\begin{figure}
\epsfig{figure=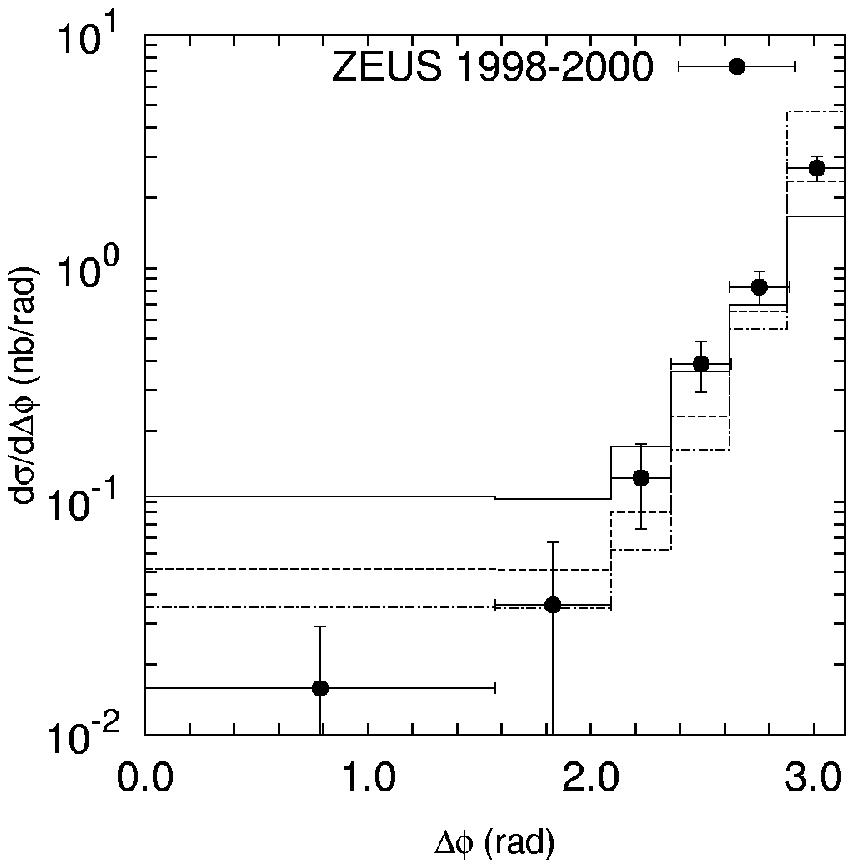, width = 22cm}
\caption{The differential cross section $d\sigma/d \Delta\phi$
for dijets with an associated $D^*$ meson with
$p_T^{D^*} > 3$~GeV, $-1.5 < \eta^{D^*} < 1.5$ in the
kinematic range $130 < W < 280$~GeV, $Q^2 < 1$~GeV$^2$, 
$-1.5 < \eta^{\rm jet} < 2.4$, $E_T^{{\rm jet}_1} > 7$~GeV, 
$E_T^{{\rm jet}_2} > 6$~GeV and $x_\gamma^{\rm obs} > 0.75$. The histograms
are the same as in Fig.~1. The experimental data are from ZEUS~[3].}
\label{fig13}
\end{figure}

\newpage

\begin{figure}
\epsfig{figure=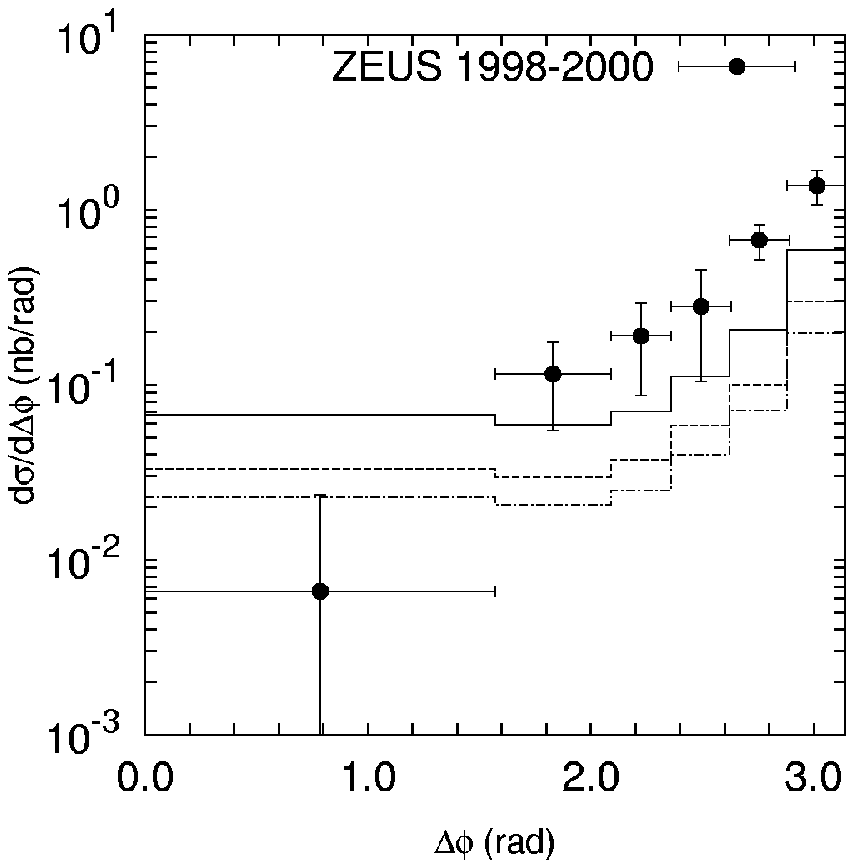, width = 22cm}
\caption{The differential cross section $d\sigma/d \Delta\phi$
for dijets with an associated $D^*$ meson with
$p_T^{D^*} > 3$~GeV, $-1.5 < \eta^{D^*} < 1.5$ in the
kinematic range $130 < W < 280$~GeV, $Q^2 < 1$~GeV$^2$, 
$-1.5 < \eta^{\rm jet} < 2.4$, $E_T^{{\rm jet}_1} > 7$~GeV, 
$E_T^{{\rm jet}_2} > 6$~GeV and $x_\gamma^{\rm obs} < 0.75$. The histograms
are the same as in Fig.~1. The experimental data are from ZEUS~[3].}
\label{fig14}
\end{figure}

\newpage

\begin{figure}
\epsfig{figure=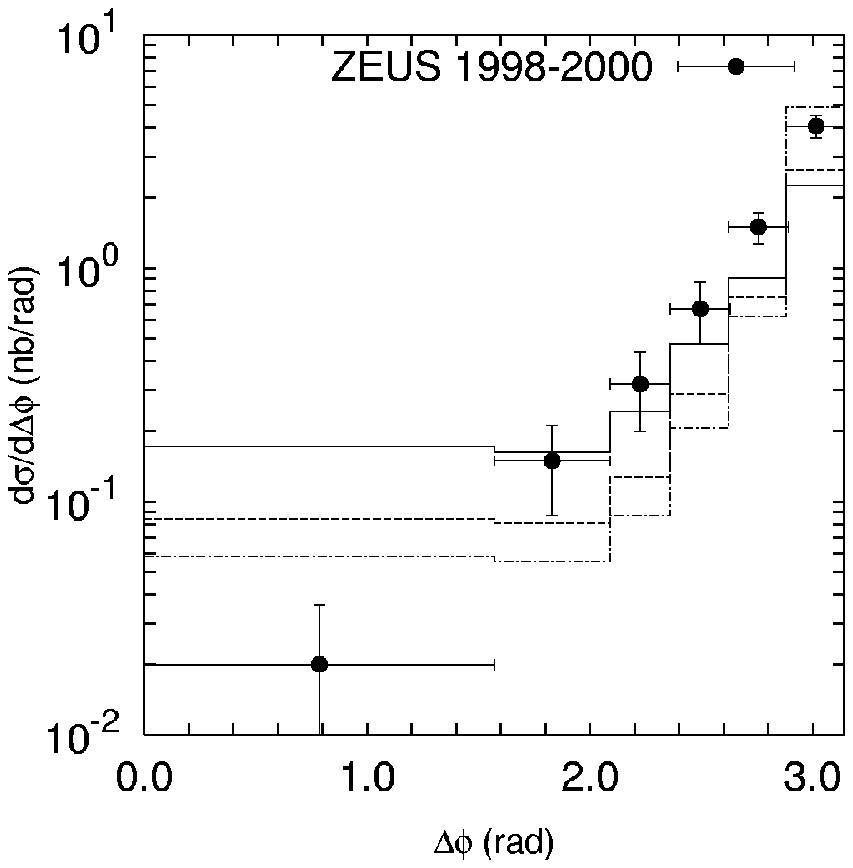, width = 22cm}
\caption{The differential cross section $d\sigma/d \Delta\phi$
for dijets with an associated $D^*$ meson with
$p_T^{D^*} > 3$~GeV, $-1.5 < \eta^{D^*} < 1.5$ in the
kinematic range $130 < W < 280$~GeV, $Q^2 < 1$~GeV$^2$, 
$-1.5 < \eta^{\rm jet} < 2.4$, $E_T^{{\rm jet}_1} > 7$~GeV 
and $E_T^{{\rm jet}_2} > 6$~GeV. The histograms
are the same as in Fig.~1. The experimental data are from ZEUS~[3].}
\label{fig15}
\end{figure}

\newpage

\begin{figure}
\epsfig{figure=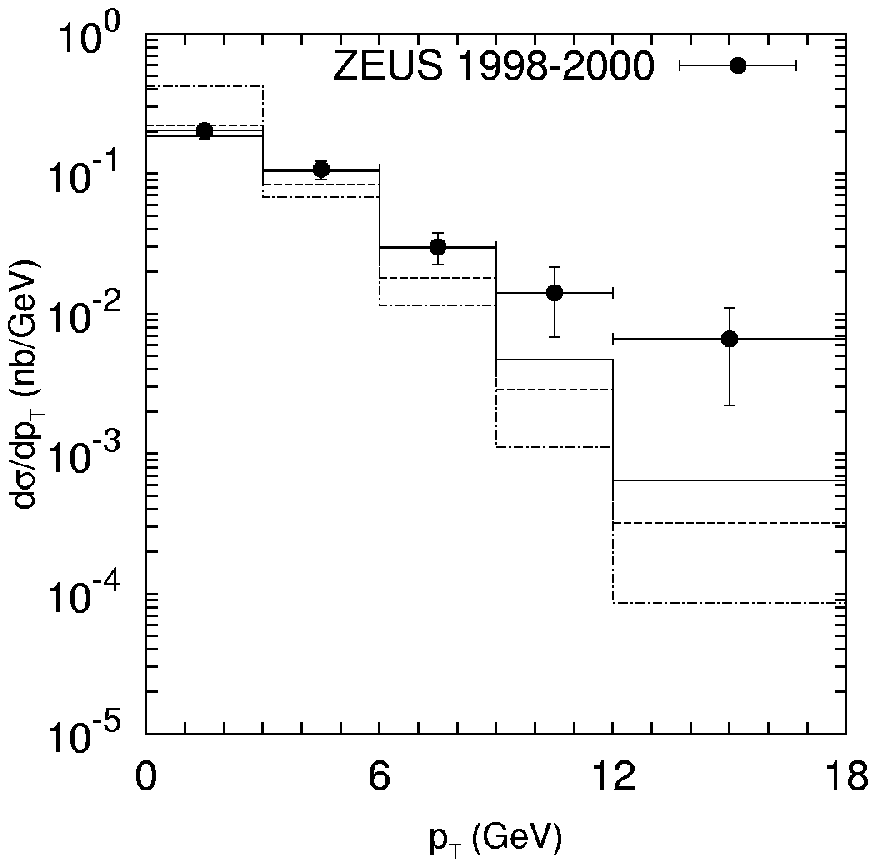, width = 22cm}
\caption{The differential cross section $d\sigma/d p_T$
for dijets with an associated $D^*$ meson with
$p_T^{D^*} > 3$~GeV, $-1.5 < \eta^{D^*} < 1.5$ in the
kinematic range $130 < W < 280$~GeV, $Q^2 < 1$~GeV$^2$, 
$-1.5 < \eta^{\rm jet} < 2.4$, $E_T^{{\rm jet}_1} > 7$~GeV, 
$E_T^{{\rm jet}_2} > 6$~GeV and $x_\gamma^{\rm obs} > 0.75$. The 
histograms
are the same as in Fig.~1. The experimental data are from ZEUS~[3].}
\label{fig16}
\end{figure}

\newpage

\begin{figure}
\epsfig{figure=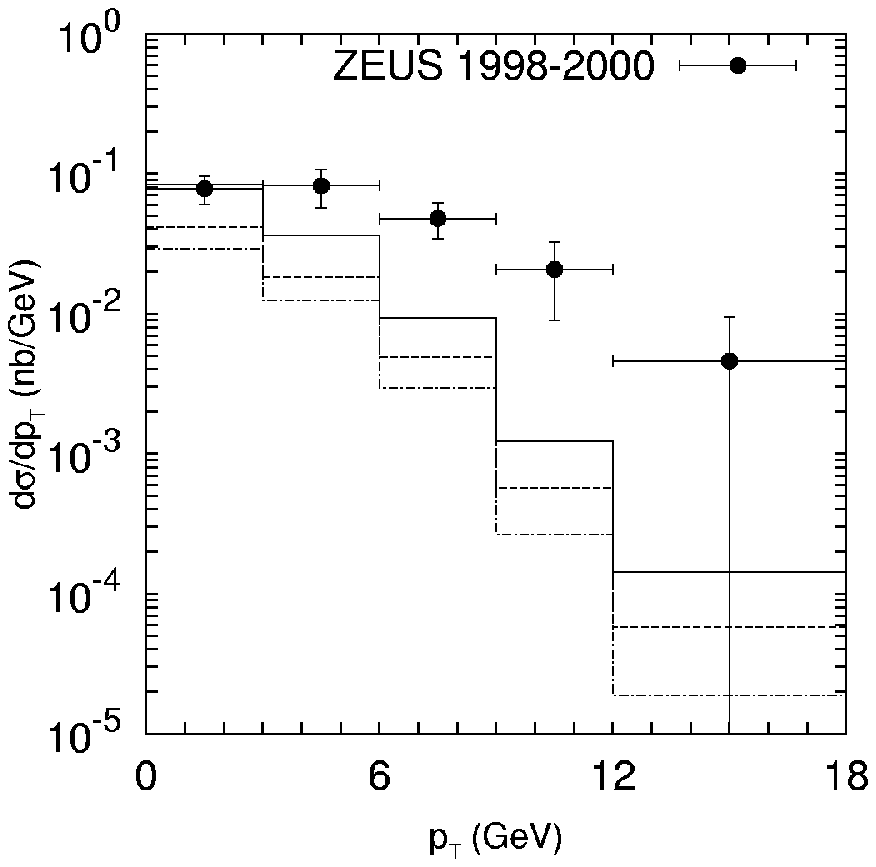, width = 22cm}
\caption{The differential cross section $d\sigma/d p_T$
for dijets with an associated $D^*$ meson with
$p_T^{D^*} > 3$~GeV, $-1.5 < \eta^{D^*} < 1.5$ in the
kinematic range $130 < W < 280$~GeV, $Q^2 < 1$~GeV$^2$, 
$-1.5 < \eta^{\rm jet} < 2.4$, $E_T^{{\rm jet}_1} > 7$~GeV, 
$E_T^{{\rm jet}_2} > 6$~GeV and $x_\gamma^{\rm obs} < 0.75$. The histograms
are the same as in Fig.~1. The experimental data are from ZEUS~[3].}
\label{fig17}
\end{figure}

\newpage

\begin{figure}
\epsfig{figure=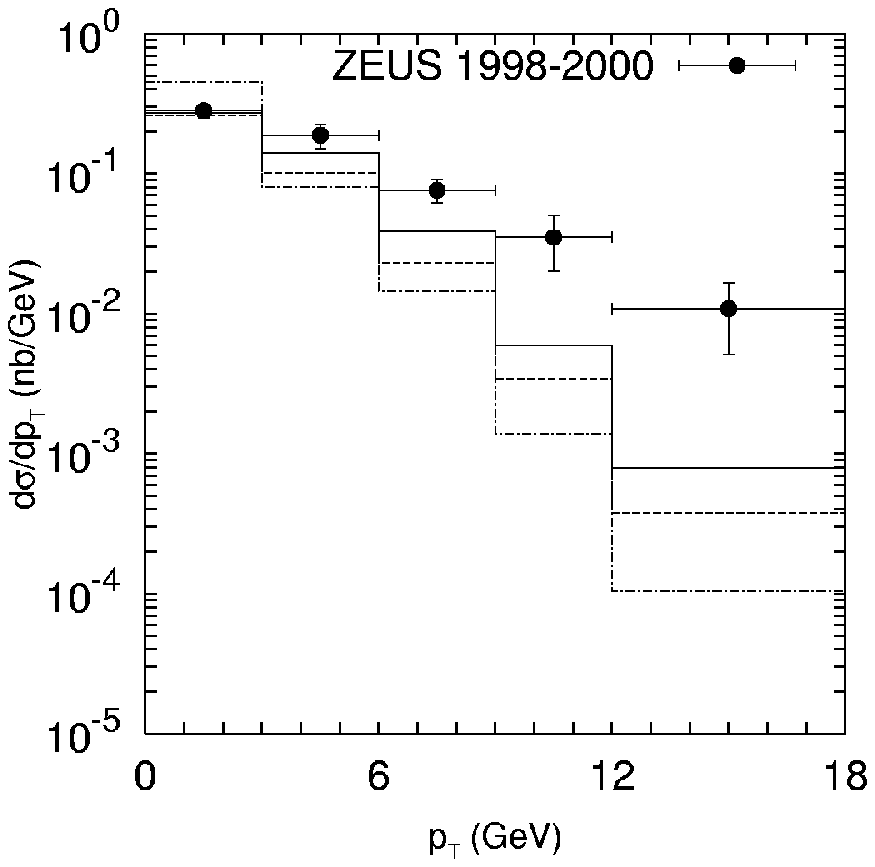, width = 22cm}
\caption{The differential cross section $d\sigma/d p_T$
for dijets with an associated $D^*$ meson with
$p_T^{D^*} > 3$~GeV, $-1.5 < \eta^{D^*} < 1.5$ in the
kinematic range $130 < W < 280$~GeV, $Q^2 < 1$~GeV$^2$, 
$-1.5 < \eta^{\rm jet} < 2.4$, $E_T^{{\rm jet}_1} > 7$~GeV 
and $E_T^{{\rm jet}_2} > 6$~GeV. The histograms
are the same as in Fig.~1. The experimental data are from ZEUS~[3].}
\label{fig18}
\end{figure}

\end{document}